\documentclass[fleqn]{article}
\usepackage{amsmath,amssymb}
\usepackage{epsfig,graphicx}

\catcode`@=11 \@addtoreset{equation}{section} \catcode`@=12

\begin{document}
\title{\vskip-1.7cm \bf  Dark energy and dark matter from nonlocal ghost-free gravity theory}
\date{}
\author{A.O.Barvinsky}
\maketitle
\hspace{-8mm} {\,\,\em Theory Department, Lebedev
Physics Institute, Leninsky Prospect 53, Moscow 119991, Russia}

\begin{abstract}
We suggest a class of generally covariant ghost-free nonlocal gravity models generating de Sitter or Anti-de Sitter background with an arbitrary value of the effective cosmological constant and featuring a mechanism of dark matter simulation. These models interpolate between the general relativistic phase on a flat spacetime background and their strongly coupled infrared (Anti)-de Sitter phase with two propagating massless graviton modes.
\end{abstract}

\maketitle

{\bf Introduction.}
It is widely recognized that one of the main challenges of modern physics is the problem of dark energy (DE) -- the mechanism which is supposed to explain observable cosmic acceleration \cite{acceleration}. Numerous efforts aimed to reconcile evidences for this phenomenon with gravity theory (\cite{quintessence,f(R),branes,massive,nonloccosm}, etc.) suffer from the fine tuning problem associated with the hierarchy of the cosmic acceleration scale vs the fundamental Planck scale. Though this problem served as a motivation to go beyond the simplest appropriate modification of general relativity (GR) -- explicit cosmological term, in this or that way it is creeping into almost all models of DE. Modulo certain exceptions \cite{Shaposhnikovetal}, most of them in fact look as a sophisticated way to incorporate into their action in addition to the Planck scale the horizon scale (whether it is a graviton mass of massive gravity \cite{massive}, multi-dimensional Planck mass in braneworld theories or the DGP scale in brane induced gravity models \cite{branes}, etc.).

To circumvent this difficulty one could adopt another, perhaps more promising, line of reasoning. If {\em a concrete fixed scale} incorporated in the model is not satisfactory, then one could look for a model that admits cosmic acceleration scenario with {\em an arbitrary scale}. Then its concrete value compatible with observations should arise dynamically by the analogue of symmetry breaking to be considered separately. Even this very unassuming approach is full of difficulties, because modified gravity models featuring this property (like unimodular gravity \cite{unimodular}, $f(R)$-gravity \cite{f(R)}, Lorentz breaking theory \cite{BlasSibiryakov}, etc.) generally violate some of its conventional symmetries and have additional degrees of freedom which might lead to ghost instabilities and make the theory inconsistent. This problem is central to numerous attempts to modify Einstein theory especially under the requirement to preserve its general covariance, and it will be a central question of this paper.

Here we suggest a nonlocal infrared modification of the metric sector of the theory, which is likely to implement the above approach. It will be based on the realization of the old idea of a scale-dependent gravitational coupling -- nonlocal Newton constant \cite{AHDDG,covnonloc,HamberWilliams} -- and will amount to the construction of the class of diffeomorphism invariant, ghost-free models compatible with the GR limit and generating the de Sitter (dS) or anti-de Sitter (AdS) background with an {\em arbitrary value} of its effective cosmological constant $\Lambda$. In addition to fine-tuning argumentation of the above type, the driving force of our approach will be the understanding of the fact that, to resolve such issues of DE as cosmic coincidence problem, this scale cannot be encoded in the fundamental or effective action of the theory (like, for instance, explicit $\Lambda$-term, massive graviton or $R+R^2/\Lambda$ models \cite{BoulwareDeser}), but rather should arise dynamically by the analogue of symmetry breaking (see, for example, \cite{Shaposhnikovetal}). Moreover, as a bonus for the construction of the ghost-free cosmic acceleration we will also get in our model a new mechanism of dark matter (DM) simulation.

{\bf Flat-space background setup.} Here we begin our search for a nonlocal modification of the Einstein theory within the concept of the effective scale-dependent gravitational constant. At a qualitative level this concept was introduced in \cite{AHDDG} as an implementation of the idea that the effective cosmological term in modern cosmology is very small not because the vacuum energy of quantum fields is so small, but rather because it gravitates too little. This degravitation is possible if the effective gravitational coupling constant depends on the momentum scale and becomes small for fields nearly homogeneous at the horizon scale. Naive replacement of the Newton constant by a nonlocal operator suggested in \cite{AHDDG} violates diffeomorphism invariance, but this procedure can be done covariantly due to the following observation \cite{covnonloc}. The Einstein action in the vicinity of flat-space background can be rewritten in the form
    \begin{equation}
    S_E=
    \frac{M_P^2}2\int dx\,g^{1/2}\left(
    -R^{\mu\nu}\frac1{\Box}G_{\mu\nu}\,
    +{\rm O}[R_{\mu\nu}^3]\right),                 \label{2.6}
    \end{equation}
where $G_{\mu\nu}=R_{\mu\nu}-\frac12g_{\mu\nu}R$ is the Einstein tensor and $1/\Box$ is the Green's function of the covariant d'Alembertian acting on a symmetric tensor\footnote{We use sign conventions for the Einstein action in the Euclidean signature spacetime and curvature tensor conventions, $R_{\mu\nu}=R^\alpha_{\;\;\mu\alpha\nu}= \partial_\alpha\Gamma^\alpha_{\nu\mu}
-\partial_\nu\Gamma^\alpha_{\alpha\mu}+...\;$.}. This expression is nothing but a generally covariant version of the quadratic part of the Einstein action in metric perturbations $h_{\mu\nu}$ on a flat-space background. When rewritten in terms of the Ricchi tensor $R_{\mu\nu}\sim \nabla\nabla h+O[h^2]$ this expression becomes nonlocal but preserves diffeomorphism invariance to all orders of its curvature expansion.

Thus, the idea of a nonlocal scale dependent Planck mass \cite{AHDDG} can be realized as the replacement of $M_P^2$ by a nonlocal operator -- a function $M^2(\Box)$ of $\Box$,
    \begin{equation}
    M_P^2 R^{\mu\nu}\frac1\Box\,G_{\mu\nu}\Rightarrow
    R^{\mu\nu}\frac{M^2(\Box)}{\Box}\,G_{\mu\nu},
    \end{equation}
which would realize this idea at least within the lowest order of the covariant curvature expansion. This modification put forward in \cite{AHDDG,covnonloc} did not, however, find interesting applications because it has left unanswered a critical question -- is this construction free of ghost instabilities for any nontrivial choice of $M^2(\Box)$? Here we try to fill up this omission and put some constraints on $M^2(\Box)$.

To begin with, if we adopt the above strategy, then the search for $M^2(\Box)$ should be encompassed by the correspondence principle. According to it nonlocal terms of the action should form a correction to the Einstein Lagrangian arising via the replacement
$R\Rightarrow R+R^{\mu\nu}F(\Box)G_{\mu\nu}$. The nonlocal form factor of this correction $F(\Box)$ should be small in the GR domain, but it must considerably modify dynamics at the DE scale. Motivated by customary spectral representations for nonlocal quantities like  $F(\Box)=\int dm^2\,\alpha(m^2)/(m^2-\Box)$ we might try the following ansatz, $F(\Box)=\alpha/(m^2-\Box)$, corresponding to the situation when the spectral density $\alpha(m^2)$ is sharply peaked around some $m^2$. As we will see, for $m^2\neq 0$ this immediately leads to a serious difficulty. Schematically the inverse propagator of the theory -- the kernel of the quadratic part of the action in metric perturbations $h_{\mu\nu}$ -- becomes $\sim -\Box+\alpha\Box^2/(m^2-\Box)$ where the squared d'Alembertian $\Box^2$ follows from four derivatives contained in the term bilinear in curvatures. Then its physical modes are given by the two roots of this expression -- the solutions of the corresponding quadratic equation $\Box=m_\pm^2$. In addition to the massless graviton with $m_-^2=0$ massive modes with $m_+^2=O(m^2)$ appear and contribute a set of ghosts which cannot be eradicated by gauge transformations (for the latter were already expended on the cancelation of ghosts in the massless sector -- longitudinal and trace components of $h_{\mu\nu}$).

Therefore, only the case of $m^2=0$ remains, and as a first step to the nonlocal gravity we will consider the action
    \begin{eqnarray}
    S=
    \frac{M^2}2\int dx\,g^{1/2}\left(-R+
    \alpha\,R^{\mu\nu}
    \frac1\Box\,G_{\mu\nu}\right).            \label{action0}
    \end{eqnarray}
On the flat-space background this theory differs little from GR provided the dimensionless parameter $\alpha$ is small, $|\alpha|\ll 1$. The upper bound on $|\alpha|$ should follow from post-Newtonian corrections in this model. The additional effect of $\alpha$ is a small renormalization of the effective Planck mass. Comparing the second term of (\ref{action0}) with (\ref{2.6}) we have in the linearized theory the following relation
    %\begin{eqnarray}
    $S=
    -\frac{M^2(1-\alpha)}2\int dx\,g^{1/2}R+\alpha O[h_{\mu\nu}^3]$,
    %\end{eqnarray}
which allows one to relate the constant $M$ to $M_P$,
    \begin{eqnarray}
    M^2=\frac{M^2_P}{1-\alpha}.  \label{M_Prenorm}
    \end{eqnarray}

{\bf Treatment of nonlocality.} At this point we have to discuss the treatment of nonlocality in (\ref{action0}). In principle, handling fundamental theories with a nonlocal action is a sophisticated and very often an open issue, because their nonlocal equations of motion demand special care in setting boundary conditions. Contrary to local field theories subject to a clear Cauchy problem setup and local canonical commutation relations, nonlocal theories can have very ambiguous rules which are critical for physical predictions. In particular, the action (\ref{action0}) above requires specification of boundary conditions for the nonlocal Green's function $1/\Box$ which will necessarily violate causality in variational equations of motion for this action.\footnote{Taken literally with any choice of boundary conditions for $1/\Box$, the action (\ref{action0}) effectively symmetrizes the kernel of this Green's function, so that nonlocal terms in equations of motion never have retarded nature and, therefore, break causality.}

To avoid these ambiguities and potential inconsistencies we will once and for all assume that our nonlocal action is not fundamental. Rather it is a certain approximation for the quantum effective action -- the generating functional of one-particle irreducible diagrams -- whose argument is the mean quantum field. This functional is necessarily nonlocal, and its nonlocality originates from quantum effects (by various mechanisms widely discussed in literature including \cite{quantum}). In this case boundary conditions for nonlocal operations are uniquely fixed by the choice of the initial (and/or final) quantum state, and manifest breakdown of causality in variational equations for this action is harmless under a proper treatment of their nonlocal terms.

To begin with, this causality breakdown does not immediately signify inconsistency in the calculation of scattering amplitudes or in-out matrix elements. These amplitudes are determined by Feynman diagrammatic technique and do not have manifest retardation properties because they are not directly physically observable. Physically observable quantities like probabilities are bilinear combinations of scattering amplitudes and can always be represented as expectation values $\langle\,{\rm in}\,|\,\hat{\cal O}\,|\,{\rm in}\,\rangle$ of certain quantum operators $\hat{\cal O}$ in the initial quantum state $|\,{\rm in}\,\rangle$. For example, the probability of transition from this state to some final state $|\,{\rm fin}\,\rangle$,
    %\begin{eqnarray}
    $P_{\,\rm in \to fin}=\langle\,{\rm in}\,|\,{\rm fin}\,\rangle\langle{\,\rm fin}\,|\,{\rm in}\,\rangle=
    \langle\,{\rm in}\,|\,\hat P_{\,\rm fin}|\,{\rm in}\,\rangle$,
    %\end{eqnarray}
is an expectation value of the projector $\hat P_{\,\rm fin}\equiv|\,{\rm fin}\,\rangle\langle{\,\rm fin}\,|$  onto this final state. In contrast to in-out matrix elements these expectation values are subject to Schwinger-Keldysh diagrammatic technique \cite{SchwKeld} which guarantees causality of $\langle\,{\rm in}\,|\,\hat{\cal O}(x)\,|\,{\rm in}\,\rangle$. This property can be formulated as a retarded response of this average to the variation of the classical external source $J(y)$ coupled to the quantum fields in terms of which the observable $\hat{\cal O}(x)$ is built,
    \begin{eqnarray}
    \frac{\delta\langle\,{\rm in}\,
    |\,\hat{\cal O}(x)\,|\,{\rm in}\,\rangle}
    {\delta J(y)}=0,      \quad x^0<y^0.     \label{causality}
    \end{eqnarray}

This property is also not manifest and turns out to be the consequence of locality and unitarity of the original fundamental field theory (achieved via a complex set of cancellations between nonlocal terms with chronological and anti-chronological boundary conditions). However, there exists a class of problems for which a retarded nature of effective equations of motion explicitly follows from their quantum effective action calculated in Euclidean spacetime \cite{beyond}. This is a statement based on Schwinger-Keldysh technique \cite{SchwKeld} that for an appropriately defined initial quantum state $|{\rm in}\rangle$ the effective equations for the mean field $g_{\mu\nu}=\langle{\rm in}|\,\hat g_{\mu\nu}|{\rm in}\rangle$ originate from the {\em Euclidean} quantum effective action $S=S_{\rm Euclidean}[g_{\mu\nu}]$ by the following procedure \cite{beyond}\footnote{We formulate this statement directly for the case of gravity theory with the expectation value of the metric field operator $\hat g_{\mu\nu}(x)$, though it is valid in a much wider context of a generic local field theory \cite{beyond}.}. Calculate nonlocal $S_{\rm Euclidean}[g_{\mu\nu}]$ and its variational derivative. In the Euclidean signature spacetime nonlocal quantities, relevant Green's functions and their variations are generally uniquely determined by their trivial (zero) boundary conditions at infinity, so that this variational derivative is unambiguous in Euclidean theory. Then make a transition to the Lorentzian signature and impose the {\em retarded} boundary conditions on the resulting nonlocal operators,
    \begin{eqnarray}
    \left.\frac{\delta S_{\rm Euclidean}}{\delta g_{\mu\nu}(x)}\right|_{\;++++\,\;
    \Rightarrow\;-+++}^{\;\rm retarded}=0.      \label{EuclidLorentz}
    \end{eqnarray}
These equations are causal ($g_{\mu\nu}(x)$ depending only on the field behavior in the past of the point $x$ in full accordance with Eq.(\ref{causality})) and satisfy all local gauge and diffeomorphism symmetries encoded in the original $S_{\rm Euclidean}[g_{\mu\nu}]$.\footnote{
A similar treatment of a nonlocal action in \cite{DeffWood} was very reservedly called the "integration by parts trick" needing justification from the Schwinger-Keldysh technique. However, this technique only provides the causality of effective equations, but does not guarantee the Euclidean-Lorentzian relation (\ref{EuclidLorentz}). The latter is based, among other things, on the choice of the $|{\rm in}\rangle$-state.
}

We will assume that our model falls into the range of validity of this procedure, which implies a particular vacuum state $|{\rm in}\rangle$ and the one-loop approximation (in which it was proven to the first order of perturbation theory in \cite{Hartle-Horowitz} and to all orders in \cite{beyond}). The extension of this range is likely to include multi-loop orders and the $|{\rm in}\rangle$-state on the (A)dS background considered below, for which this state apparently coincides with the Euclidean Bunch-Davies vacuum.

Thus, the action (\ref{action0}) is understood as the Euclidean one (this explains our sign choice in the Einstein term) with zero boundary conditions for $1/\Box$ at infinity. It can be localized in terms of the auxiliary tensor field subject to the same Dirichlet boundary conditions\footnote{This field formally carries ghosts, but this does not indicate physical instability because it never exists as a free field in the external lines of Feynman graphs. The actual particle content of the theory is determined in terms of the original metric field $g_{\mu\nu}$ and indeed turns out to be ghost-free on the flat-space background, because the quadratic part of the action coincides with the Einstein's one. A similar mechanism excluding ghosts by boundary conditions was recently used in the conformal gravity model of \cite{Maldacena}.}, and in the resulting local representation directly applied to the FRW cosmology. This shows that close to a certain moment $t_0$ corresponding to the present epoch the model easily yields a (quasi) de Sitter point of the cosmological evolution \cite{nonlocPRD}, its Hubble factor $H=\dot a/a$ and the equation of state parameter $w=-1-2\dot H/3H^2$ satisfying the relations $w(t_0)=-1$, $\dot w(t_0)=O(1)\times H(t_0)<0$, which make the model qualitatively compatible with the observable DE data. These preliminary estimates could have served as a starting point for a quantitative comparison with the DE scenario. However, a formal application of (\ref{action0}) to the FRW setup disregards nontrivial boundary conditions in cosmology. To see this, note that on the de Sitter background (which is a zeroth-order approximation for the cosmic acceleration scenario) the Ricci curvature $R_{\mu\nu}=\Lambda g_{\mu\nu}$ is covariantly constant, and the nonlocal part of (\ref{action0}) is divergent, because $g_{\mu\nu}$ is a zero eigenvector of $\Box$. This means that the action (\ref{action0}) should be modified to circumvent this difficulty.

{\bf Nonlocal gravity with a stable (A)dS background.} We will regulate the action (\ref{action0}) by adding to the generally covariant $\Box$ the matrix-valued potential term built of a generic combination of tensor structures linear in the curvature,
    \begin{eqnarray}
    &&S=\frac{M^2}2\int dx\,g^{1/2}\,\left(-R+
    \alpha\,R^{\mu\nu}
    \frac1{\Box+\hat P}\,G_{\mu\nu}
    \right),\;\;\;\;                        \label{action}\\
    &&\hat P\equiv P_{\alpha\beta}^{\;\;\;\mu\nu}
    =a R_{(\alpha\;\;\beta)}^{\;\;\,(\mu\;\;\,\nu)}
    +b \big(g_{\alpha\beta}R^{\mu\nu}
    +g^{\mu\nu}R_{\alpha\beta}\big)
    %\nonumber\\
    %&&\qquad\qquad\qquad
    +c R^{(\mu}_{(\alpha}\delta^{\nu)}_{\beta)}
    +d R\,g_{\alpha\beta}g^{\mu\nu}
    +e R \delta^{\mu\nu}_{\alpha\beta}.     \label{potential}
    \end{eqnarray}
Here we use the condensed notation for the Green's function of the operator $\Box+\hat P\equiv\Box\,\delta_{\alpha\beta}^{\;\;\;\mu\nu}
+P_{\alpha\beta}^{\;\;\;\mu\nu}$, acting on a symmetric tensor field as
    \begin{eqnarray}
    \frac1{\Box+\hat P}\,G_{\mu\nu}\equiv
    \Big[\,\frac1{\Box+\hat P}\,\Big]_{\mu\nu}^{\alpha\beta}G_{\alpha\beta}
    \end{eqnarray}
and $a$, $b$, $c$, $d$ and $e$ represent arbitrary parameters to be restricted by the requirement of a stable (A)dS solution in the model. Of course, such a modification of the original action (\ref{action0}) leaves its linear approximation on a flat background intact, because it deals with $O[h_{\mu\nu}^3]$-terms.

Now the Green's function $1/(\Box+\hat P)$ acting on the Einstein and Ricci tensors in (\ref{action}) is well defined even for the (A)dS background with the covariantly constant $R_{\mu\nu}=\Lambda g_{\mu\nu}$ and $R_{\alpha\mu\beta\nu}=\frac{\Lambda}3 (g_{\alpha\beta} g_{\mu\nu}-g_{\alpha\nu}g_{\beta\mu})$, for which
    \begin{eqnarray}
    &&P_{\alpha\beta}^{\;\;\mu\nu}
    =\frac{A+4B}4\,\Lambda g_{\alpha\beta}g^{\mu\nu}
    -C\Lambda\left(\delta^{\mu\nu}_{\alpha\beta}-\frac14\,
    g_{\alpha\beta}g^{\mu\nu}\!\right),\\
    &&A=a+4\,b+c,\,\,\,\,B=b+4\,d+e,       \label{A}\\
    &&C=\frac{a}3-c-4e,                    \label{C}
    \end{eqnarray}
so that $\hat P\,g_{\mu\nu}\equiv P_{\mu\nu}^{\;\;\;\;\alpha\beta}\,g_{\alpha\beta}
    =(A+4B)\,\Lambda g_{\mu\nu}$.

Let us show that under a certain restriction on parameters of $\hat P$ the model (\ref{action}) has (A)dS solution with an {\em arbitrary} value of the cosmological constant $\Lambda$. Indeed, introduce the local conformal variation with the parameter $\delta\sigma=\delta\sigma(x)$,
    %\begin{eqnarray}
    $\delta_\sigma=\int d^4x\,\delta\sigma\,g_{\alpha\beta}\,\delta/\delta g_{\alpha\beta}$.
    %\end{eqnarray}
It operates on various quantities in (\ref{action}) according to their conformal weights, $\delta_\sigma g_{\mu\nu}=\delta\sigma\,g_{\mu\nu}$,
    $\delta_\sigma R_{\mu\nu}=O(\nabla)$, $\delta_\sigma R=-\delta\sigma\, R+O(\nabla)$, $\delta_\sigma\hat P=-\delta\sigma\,\hat P+O(\nabla)$, $\delta_\sigma\Box=O(\nabla)$,
modulo the derivatives $O(\nabla)$ acting on $\delta\sigma(x)$ and these quantities themselves. Therefore, the conformal variation of (\ref{action}) {\em on the (A)dS background} reads
    \begin{eqnarray}
    &&\left.\frac{\delta S}{\delta\sigma}\,
    \right|_{\;\rm (A)dS}=
    \frac{M^2}2 g^{1/2}\left(-R+\alpha\,
    R^{\alpha\beta}\hat P^{-1}\,
    G_{\alpha\beta}\,\right)
    =-2M^2\Lambda
    \left(1+\frac\alpha{A+4B}\right)g^{1/2}.      \label{13}
    \end{eqnarray}
On this background the full metric variational derivative expresses via its trace part given by this variation,
    %\begin{eqnarray}
    $\delta S/\delta g_{\mu\nu}\,|_{\rm (A)dS}=\frac14\,g^{\mu\nu}(\delta S/\delta\sigma)\,|_{\rm (A)dS}$.
    %\end{eqnarray}
Therefore, the equation of motion $\delta S/\delta g_{\mu\nu}\,|_{\rm (A)dS}=0$, holds with an arbitrary value of $\Lambda$ when
    \begin{eqnarray}
    \alpha=-A-4B.    \label{relation}
    \end{eqnarray}

Note that the existence of the (A)dS solution with an arbitrary $\Lambda$ is neither the result of the local Weyl invariance of the theory, nor even its global scale invariance. Rather this is a corollary of the relation (\ref{relation}) which, in particular, guarantees the vanishing on-shell value of the action $ S\,|_{\rm (A)dS}=0$. Thus, this solution is another vacuum -- a direct analogue of the flat-space one.

Another remarkable consequence of Eq.(\ref{relation}) is the stability of the (A)dS solution against ghost and tachyon excitations. In principle, the hope to eradicate ghosts and tachyons from the quadratic part of the action $S_{(2)}$ on the (A)dS background is based on the observation that in the DeWitt gauge,
    \begin{eqnarray}
    \chi^\mu\equiv\nabla_\nu
    h^{\mu\nu}-\frac12\,\nabla^\mu h=0,          \label{DWgauge}
    \end{eqnarray}
$S_{(2)}$ contains only two contractions $h_{\mu\nu}^2$ and $h^2$ ($h\equiv g^{\mu\nu}h_{\mu\nu}$), their nonlocal parts being given by $h^{\mu\nu}(\Box+\hat P)^{-1}h_{\mu\nu}$ and $h(\Box-\alpha\Lambda)^{-1}h$.\footnote{The form of the Green's function in the trace sector follows from the equation $(\Box+\hat P)g_{\mu\nu}h=g_{\mu\nu}(\Box-\alpha\Lambda)h$ also based on (\ref{relation}).} As in the discussion above, the ghosts necessarily appear if these nonlocalities are nonvanishing in $S_{(2)}$, because the dispersion equation for $\Box$ becomes quadratic and generates doubled set of physical modes with $\Box=m_\pm^2$ ($\hat P$ playing the role of nonvanishing $m^2$ above). Therefore it is a priori possible to cancel these two nonlocalities and provide the right signs of the remaining local terms by an appropriate choice of five parameters in (\ref{potential}).

Curious fact is that in the DeWitt gauge $S_{(2)}$ depends only on the traceless part of $h_{\mu\nu}$, $\tilde h_{\mu\nu}=h_{\mu\nu}-\frac14\,g_{\mu\nu}h$, and is very simple
    \begin{eqnarray}
    &&S_{(2)}\!=-\frac1{64\pi G_{\rm eff}}
    \int d^4x g^{1/2}\left\{\tilde h^{\mu\nu}\Box\,
    \tilde h_{\mu\nu}+\Big(C-\frac43\Big)\Lambda
    \tilde h_{\mu\nu}^2\right.
    %\nonumber\\
    %&&\qquad\qquad\qquad\qquad
    +\Lambda^2\Big(C-\frac23\Big)^2
    \left.\tilde h^{\mu\nu}\frac1{\Box+\hat P}\,\tilde h_{\mu\nu}
    \right\}, \label{s_2}\\
    &&G_{\rm eff}=
    \frac{\alpha(1-\alpha)}{8B(2B+\alpha)}\,G_N,                                 \label{Geff}
    \end{eqnarray}
where $G_{\rm eff}\equiv 1/8\pi M_{\rm eff}^2$ is the effective gravitational constant vs the Newton constant $G_N=1/8\pi M_P^2$. This is the main technical result of this paper (the detailed derivation of which will be reported in \cite{nonlocPRD}), and it leads to the single equation for the constant (\ref{C}),
    \begin{equation}
    C\equiv\frac{a}3-c-4e=\frac23,   \label{Crelation}
    \end{equation}
and the positivity requirement for $G_{\rm eff}$. This requirement selects the range of the parameter $B$, $B<-\alpha/2$ and $B>0$ for a positive $\alpha$, and even more interesting compact range of $B$ for a negative $\alpha$,
    \begin{eqnarray}
    &&0<B<-\frac\alpha2,\,\,\,\,\,\alpha<0.   \label{negativealpha}
    \end{eqnarray}

The action without gauge-fixing can be obtained from (\ref{s_2}) by
representing $h_{\mu\nu}$ in the DeWitt gauge as the projection of the non-gauged field,
    $h_{\mu\nu}|_{\chi^\alpha=0}=h_{\mu\nu}
    -2\,\big[\nabla_{(\mu}\delta^\alpha_{\nu)}/
    (\Box+\Lambda)\big]\,\chi_\alpha$.
The result reads
    \begin{eqnarray}
    &&S_{(2)}=\frac1{16\pi G_{\rm eff}}\int d^4x\,g^{1/2}\left\{\,\frac14\,h^{\mu\nu}
    \left(-\Box+\frac23\,\Lambda\right)\, h_{\mu\nu}
    \right.
    %\nonumber\\
    %&&\qquad\quad\quad
    -\frac18\,h\left(-\Box-\frac23\,\Lambda\right)\,h
    -\frac12\,\chi_\mu^2\nonumber\\
    &&\qquad\qquad\qquad\qquad
    %\nonumber\\
    %&&\qquad\quad\quad
    \left.-\frac14\,R_{(1)}\frac1{\Box+2\Lambda}
    R_{(1)}\right\},   \label{ngaction}
    \end{eqnarray}
where, quite interestingly, the first three terms coincide with the quadratic part on the (A)dS background of the Einstein-Hilbert action with the $\Lambda$-term and  $R_{(1)}=\nabla_\mu\chi^\mu-\frac12(\Box+2\Lambda)h$ is the linearized Ricci scalar on the same background.

The trace of the variational equation for (\ref{ngaction}), according to (\ref{EuclidLorentz}), gives the homogeneous equation for $R_{(1)}$ with a retarded nonlocality, which is equivalent to the local initial Cauchy problem $\Box R_{(1)}=0$ with zero initial data in the remote past \cite{nonlocPRD}. Therefore $R_{(1)}(x)\equiv 0$, whence in the DeWitt gauge $\chi^\mu=0$ we have $(\Box+2\Lambda)\,h=0$. Then the residual gauge transformations $\Delta^f h_{\mu\nu}=2\nabla_{(\mu} f_{\nu)}$ with the parameter $f_\mu$ satisfying the equation $(\Box+\Lambda)f_\mu=0$ can be used to select two polarizations -- non-ghost physical modes. In particular, the  boundary conditions for $h$ can be nullified, so that $h$ identically vanishes and makes in view of the DeWitt gauge the propagating free modes transverse and traceless as in the Einstein theory with the $\Lambda$-term.

In the presence of matter sources with a stress tensor $T_{\mu\nu}$  of a compact support the causal effective equations for retarded gravitational potentials become local in the DeWitt gauge,
    \begin{eqnarray}
    \left(-\Box+\frac23\Lambda\right) h_{\mu\nu}+\frac12\nabla_\mu\nabla_\nu h-\frac\Lambda6 g_{\mu\nu}h=
    \frac2{M_{\rm eff}^2}T_{\mu\nu}.         \label{coupledmode}
    \end{eqnarray}
Modulo the gauge transformation their solution takes the following form -- the result of a careful commutation of covariant derivatives with $(-\Box+\frac23\,\Lambda)^{-1}$,
    \begin{equation}
    h_{\mu\nu}
    =\frac{8\pi G_{\rm eff}}{-\Box+\frac23\Lambda}\,\left(T_{\mu\nu}
    +g_{\mu\nu}\,
    \frac{\Box-2\Lambda}{\Box+2\Lambda}\,
    \frac{\Lambda}{3\Box}\,T\right).     \label{matpot}
    \end{equation}
The tensor structure here differs from the GR analog $T_{\mu\nu}-\frac12g_{\mu\nu}T$, which for non-relativistic sources gives $O(1)$ correction. What is much more interesting, it yields an unexpected bonus in the form of the dark matter simulation -- $1/|\alpha|$-amplification of the gravitational attraction due to the replacement of the Newton gravitational constant $G_N$ by (\ref{Geff}), $G_{\rm eff}\sim G_N/|\alpha|$ with $|\alpha|\ll 1$. This is possible when $|B|\sim\alpha$ for a positive $\alpha$ and necessarily happens in the case (\ref{negativealpha}) of a negative $\alpha$, because the factor $\alpha/8B(2B+\alpha)\geq 1/|\alpha|$ and
    \begin{eqnarray}
    G_{\rm eff}\geq\frac{1-\alpha}{|\alpha|}\,G_N\gg G_N. \label{10}
    \end{eqnarray}
On the other hand, with $|B|\sim\sqrt\alpha$ in the case of a positive $\alpha$ both Newton and effective gravitational coupling constants can be of the same order of magnitude, $G_N/G_{\rm eff}=O(1)$ even for $\alpha\ll 1$, which together with (\ref{10}) leaves a large window for a possible strength of DM attraction relative to GR behavior.

{\bf Conclusions.} The theory (\ref{action}) has two phases. For short distances corresponding to the range of wavelengths with $\nabla\nabla\sim\Box\gg R$ this is a GR phase on the zero curvature background with small $O(\alpha)\times R/\Box$ corrections of higher orders in spacetime curvature (collectively denoted by $R$). This regime applies to galactic, Solar system and other small scale phenomena and is likely to pass all general relativistic tests for a sufficiently small $\alpha$.\footnote{Disturbing property of perturbation theory in this phase could be the presence of poles $\sim(1/\Box)R$ in the vertices of (\ref{action}) which could contribute too strong to graviton scattering. But these contributions vanish on shell $R_{\mu\nu}=0$ provided, perhaps, that the Riemann term is forbidden in (\ref{potential}), $a=0$. This question requires further studies.}

Another phase of the theory corresponds to the infrared wavelengths range $\nabla\nabla\ll R$ in which a stable (A)dS background exists and the modified gravitational potential of matter sources is given by Eq.(\ref{matpot}). This equation is valid for the perturbation range $|\delta R^\mu_\nu|\sim|\nabla\nabla h^\mu_\nu|\ll \Lambda$ and $|h^\mu_\nu|\ll 1$ equivalent to very small matter densities $|T^\mu_\nu|\ll M_P^2\Lambda$ characteristic of galaxy, galaxy cluster and horizon scales for which DE and DM modification of gravity theory becomes important. Thus, nonlocal gravity interpolates between GR theory and its strongly coupled infrared modification which is likely to generate a stable ghost-free stage of cosmic acceleration and, perhaps, even simulate the DM effect on rotation curves.

Of course, prospective nature of this model should not be exaggerated. Mainly, it challenges the ghost and fine-tuning problems in infrared modifications of Einstein theory, designed to simulate DE and DM mechanisms in the {\em modern} Universe.
Without a proper extension this model most likely fails to describe early inflationary cosmology. In particular, it fails to generate the (quasi)de Sitter solution by the external cosmological constant $\lambda$ (vacuum energy of matter fields), because the relation (\ref{relation}) in this case becomes broken by a new scale $\lambda$, $\alpha=-(A+4B)(1+\lambda/\Lambda)$. Of course, this relation can be considered as the equation for $\Lambda$ in terms of $\lambda$ and the stability analysis can be repeated, but the spirit of our approach avoiding the fine-tuning problem becomes lost. This property works in favor of a supersymmetric matter content of the Universe with a vanishing $\lambda$.

In contrast to a homogeneous vacuum energy source $T_{\mu\nu}\propto \lambda\, g_{\mu\nu}$, matter sources with a compact support present no difficulty and generate a stable perturbation theory on the (A)dS background, starting with the linearized potentials (\ref{matpot}). This perturbation theory is applicable in the present day cosmology, and it should undergo tests on consistency of post-Newtonian corrections, on the magnitude of its DM phenomenon and the effect of the nonlocal trace term in (\ref{matpot}). Moreover, the mechanism should be found, by which the model picks up a concrete scale $|T^\mu_\nu|\sim M_P^2\Lambda$ of a crossover from the GR regime to cosmic acceleration -- necessary element in realistic cosmology.\footnote{Note that the existence of (A)dS solution with an arbitrary $\Lambda$ is possible due to the fact that the purely gravitational action (\ref{action}) transforms homogeneously under the global rescaling, $g_{\mu\nu}\to\lambda g_{\mu\nu}$, $S[\lambda g_{\mu\nu}]=\lambda S[g_{\mu\nu}]$, so the crossover mechanism can be based on the fact that matter fields break this property, cf. \cite{Shaposhnikovetal}.} The last but not the least is the justification of the choice of (\ref{action}) as an approximation for the effective action coming from some fundamental quantum gravity theory. These and other open issues that go beyond this paper will be discussed in \cite{nonlocPRD}.

In conclusion we mention that serendipity of ghost-free nonlocal gravity models (\ref{action}) satisfying the relations (\ref{relation}) and (\ref{Crelation}) might not be exhausted by applications in cosmology. In particular, without the Riemann term in (\ref{potential}) ($a=0$) they admit generic (not maximally symmetric) Einstein space solutions, $R_{\mu\nu}=\Lambda g_{\mu\nu}$, also with an {\em arbitrary} $\Lambda$, which might have implications for zero entropy black holes \cite{Solodukhin} and be an alternative to the conformal gravity model of \cite{Maldacena}, whereas with a negative $\Lambda$ they become a new testing ground for AdS/CFT correspondence perhaps promising other exciting consequences.

\section*{Acknowledgements}
The author strongly benefitted from thought-provoking criticism of G. Dvali and fruitful discussions with O. Andreev, S. Hofmann, V. Mukhanov, I. Sachs, M. Shaposhnikov and S. Solodukhin.  This work was supported by the Humboldt Foundation at the Physics Department of the Ludwig-Maximilians University in Munich and by the RFBR grant No 11-02-00512.

\end{document}